\documentstyle[12pt]{article}

\title{\bf On the odderon intercept in the perturbative QCD}
\author{M.A.Braun\\ Department of High Energy Physics,
University of St. Petersburg,\\ 198904 St. Petersburg, Russia.}
\date{}
\def\beq{\begin{equation}}
\def\eeq{\end{equation}}
\def\noi{\noindent}

\begin{document}
\maketitle
\medskip
\noi{\bf Abstract.}

The odderon intercept is calculated directly using the wave function
recently constructed by R.A.Janik and J.Wosiek. The results confirm
their reported value.  It is also found that their solution for
$q_3=0$ does not satisfy the Bose-symmetry requirements. Introduction of
terms into the trial wave function with an asymptotical behavior similar
to the Janik-Wosiek wave functions does not seem to improve variational
estimates significantly.  The diffusion parameter is found to be of the
order $0.6$.
\vspace{6.5cm}

\noi{\Large\bf SPbU-IP- 1998/8}
\newpage
\section{Introduction. Basic equations.}
{\bf 1.} Recently R.A.Janik and J. Wosiek (JW)
published a report on the solution
of the odderon problem in the perturbative QCD [1]. They derived the odderon
intercept from the spectrum of the integral of motion $q_3$
introduced by L.N.Lipatov for the
three-Reggeon system [2], combined with their earlier solution of the
appropriate Baxter equation [3]. If one relates the odderon intercept
to the odderon "energy" per pair of Reggeons $\epsilon$ as
\beq
\alpha_O(0)=1-(3\alpha_s/2\pi)\epsilon
\eeq
the result of JW for the ground state is
\beq
\epsilon=0.16478...
\eeq
Thus they confirm our old conclusion that the odderon intercept lies below
unity [3].  Their exact value for it is somewhat higher than
obtained in variational calculations, which gave larger value for
$\epsilon$: 0.29 [4] and 0.223 [5,6].

In view of a somewhat indirect way of finding the odderon energy,
JW suggested verifying their result by means of
direct calculations of the energy with their wave function, using the
technical machinery developed in the variational approach.
Also, to prove that their result indeed gives the minimal value for the
energy, it is desirable to repeat  variational calculation using more
general trial functions of the form suggested by the JW
solution. This short note presents results of these calculations.
\vspace{0.8cm}

{\bf 2.} The odderon energy can be sought as a ratio
\beq
\epsilon=E/D
\eeq
where  $E$ and $D$ are energy and normalization functionals, quadratic
in the odderon wave function $Z(r,\phi)$ [5]. Explicitly
\beq
E=\sum_{n=-\infty}^{\infty}\int_{-\infty}^{\infty}d\nu\epsilon_{n}(\nu)
|\alpha_{n}(\nu)|^{2}
\eeq
Here
\beq
\epsilon_{n}(\nu)=2\ {\mbox Re}\,\left(\psi(\frac{1+|n|}{2}+i\nu)
-\psi(1)\right)
\eeq
$\alpha_n(\nu)$ is esentially a double Fourier transform of $Z(r,\phi)$:
\[
\alpha_{n}(\nu)=\int_{0}^{\infty}dr r^{-2-2i\nu}\int_{0}^{2\pi}
d\phi e^{-in\phi}\]\beq
\left(i\nu+\frac{n+1}{2}+re^{i\phi}(h-i\nu-\frac{n-1}{2})\right)
(i\nu-\frac{n-1}{2})(-\tilde{h}+i\nu-\frac{n-1}{2})Z(r,\phi)
\eeq
where
$ h $ and
$ \tilde{h} $ are the two conformal weights, which are taken
to be equal to 1/2 for the ground state.
$D$ is obtained substituting  $\epsilon_n(\nu)$ by unity in (1).

The odderon wave function $Z$ can be considered as a function of
$z=r\exp i\phi$ and its conjugate. It has a form
\beq
Z(z,z^*)=|z(1-z)|^{1/3}\Phi(z,z^*)
\eeq
Bose symmetry requires invariance of $\Phi$ under substitutions
\beq z\rightarrow 1-z,\ \ z\rightarrow 1/z
\eeq
Function $\Phi$ was explicitly constructed by JW (in the
form of a power series).

Once $\Phi$ is known, calculation of $\epsilon$ reduces to, first,
performing the double Fourier transform (6) and, second, doing the
integration over $\nu$ and summation over $n$ in (4).

All the difficulty in the calculation resides at the first stage.
Actually one has to go to quite high values of $n$ and $\nu$ to get
reasonable accuracy. The Fourier transform to such high values
of $n$ and $\nu$ requires much effort. It also requires a very precise
knowledge of the wave function $\Phi$

\section{Odderon ground state.}
{\bf 1.}
Because of the latter requirement our first step was to
obtain the odderon wave function with a higher precision than
reported in [1].
To this end we set up a program which essentially repeats the procedure
employed in [1] and allows to reduce discontinuities of $\Phi$
calculated in different variables to values of the order $10^{-9}$.
The only difference with [1] is that we use the basic functions $u_i$,
$i=1,2,3$, multiplied by certain factors to make them real at
points where the transfer matrices are calculated. This substantially
facilitates achieving the desired accuracy.

 At this step  for the integral of motion $q_3$ and wave
function parameters $\alpha$, $\beta$ and $\gamma$ we obtained the following
values (precision $10^{-9}$)
\beq
iq_3=0.205257506,\ \ 
\alpha=0.709605410,\ \ 
\beta=-0.689380668,\ \ 
\gamma=0.145651837
\eeq

However even with these high precision values direct calculation of $\Phi$
in one of the three sets of variables $z$, $1-1/z$ or $1/(1-z)$ fails in the
vicinity of the point $z_0=\exp i\pi/3$ where none of the series
converges absolutely. In spite of the fact that $z_0$ is
only a single point in the $r,\phi$ plane, this makes it practically
impossible  to
calculate the double Fourier transform for $|n|>10$ and/or $|\nu |>5$.
To overcome this difficulty we had to redevelop the function
$\Phi$ around the intermediate point $(1/2)z_0$. With this redevelopment
100 terms in the series proved to be sufficient to obtain reliable results.

The Fourier transform itself was performed by interpolating $Z$
quadratically
on the $r,\phi$ grid and doing the integrals in $r$ and $\phi$
analytically. Reasonable results are obtained already with a 160 x 160
grid. We however also used  320 x 320 and 640 x 640 grids to
analyse the precision achieved at this step.

Even with a 640 x 640 grid the numerical Fourier transform becomes
unreliable for $|n|>30$ and/or $|\nu |>15$. For such high values of
$|n|$ and $|\nu|$ we used asymptotic formulas for the Fourier transform,
which can easily be obtained from the expansion of $\Phi$ around $z=1$.
Our final cutoffs were chosen to be $|n|<300$ and $|\nu|<150.$,
which proved to be quite sufficient for the determination of $\epsilon$
with a precision  $0.001$
\vspace{0.8cm}

{\bf 2.} Results of our numerical calculation of $D$ and $E$ in
the region $|n|<30$ and $|\nu|<15$ using an $N$ x $N$ grid in the
$r,\phi$ plane are presented in the Table 1. for $N=160,320$ and $640$

\begin{center}
{\bf Table 1. $D$ and $E$ for the ground state}\vspace{0.5cm}

\begin{tabular}{|r|c|c|}\hline
N& D& E\\\hline
160&1.642162&0.255480\\\hline
320&1.642085&0.254852\\\hline
640&1.642056&0.254632\\\hline
\end{tabular}
\end{center}

To these values one has to add the contributions from the asymptotic
region described in the preceding section. They are
\beq
\Delta D=0.002398,\ \ \Delta E=0.018451
\eeq
Taking the results at $N=640$ as the most accurate ones we finally have
\beq
D=1.644454,\ \ E=0.273083,\ \ \epsilon=0.1660
\eeq
Thus our result for $\epsilon$ coincides with the value found in [1]
up to 0.001. With all the difficulties involved in the numerical
calculations we  consider this agreement  quite satisfactory.
So direct calculation of the odderon energy confirms the result found
by JW.\vspace{0.8 cm}

{\bf 3.} Our previous variational calculations 
gave  very small change in energy as more analytic terms were added.
To analyse the reason of the about 30\% improvement given by the
JW wave function,
we tried to extend our variational calculations to
more general trial functions as compared to [5,6], whose form is
suggested
 by the latter function.
 At $q_3\neq 0$ it possesses an asymptotical behavior
 at $z\rightarrow 0$ of the same sort as the trial wave function
 introduced by P.Gauron, L.N.Lipatov and B.Nicolescu in [5], except
 of a term which behaves as $r^{5/3}\cos 2\phi$. To investigate
 the importance of this behaviour we introduced a term into the
 trial wave function
 \beq
           ca(z,z^*)^{-1/6}b(z,z^*)^2
 \eeq
 Here $a$ is the argument in the trial wave function of [5]:
 \beq
 a=x/y,\ \ x=|z|^2|z_1|^2,\ \ y=(1+|z|^2)(1+|z_1|^2)(|z|^2+|z_1|^2)
 \eeq
 where $z_1=1-z$. The argument $b$ is
 \beq
 b=w/y,\ \ w=(1-|z|^2)(1-|z_1|^2)(|z|^2-|z_1|^2)
\eeq
It is invariant under $z\rightarrow 1-z$ and changes sign under
$z\rightarrow 1/z$, so that $b$ is invariant.

At $q_3=0$ a term appears in the solution $\Phi$ which is unique in the
$r,\phi$ plane and blows up as $r^{-1/3}$ at $r\rightarrow 0$.
Inspection of the functionals $D$ and $E$ shows however that it is
admissible in spite of the apparent singularity at small $r$ in (6).
As mentioned, we could not satisfy the Bose symmetry requirements with
this solution. Nevertheless we tried to estimate its possible
significance and so included a term proportional to $a^{-1/6}$
into the trial function.

Our final trial function thus included 5 terms, three old ones of the
same form as in [5,6] and two new ones described above.

Calculations with this generalized trial function
showed first of all that the  term with $A^{-1/6}$ presents
difficulties for numerical integration, of the same sort that
we encountered in studying the excited odderon, only much worse.
In fact, with this term added, we could only calculate the double
Fourier transform reliably for $|n|<8$, $|\nu|<5.$ On the other hand,
the term with $b$ lead to no difficulties whatsoever. However in
both cases one finds that
the new added terms bear no influence on the value of energy.
At the minimum value for the functional $E$ the coefficients before
them turn out to be quite small and the value itself is only a few
percent lower than without the new terms.
So our conclusion is that simple addition of new terms into the
trial wave functions even with two independent arguments $a$ and $b$
described above does not improve variational estimates. It is a
factorized form of the Janik-Wosiek wave function which allows to make energy
substantially lower. These calculations also make us believe that
the JW function indeed belongs to the odderon ground state.

\section{Other eigenvalues of $q_3$}

{\bf 1.} We have also tried to check the result of [1] for the excited
state with the next higher value of $iq_3$. Unfortunately in this
case calculations proved to be still more difficult and we could not
arrive at  a result of a convincing accuracy.

Our precise calculations  of the wave function gave for this state:
\beq
iq_3=2.343921063,\ \ 
\alpha=0.391855163,\ \ 
\beta=-0.0533712012,\ \ 
\gamma=0.918477570
\eeq

Numerical calculation of $D$ and $E$ in the region
$|n|<15$, $|\nu|<15.$ gave results presented in Table 2.

\begin{center}
{\bf Table 2. $D$ and $E$ for the state with $iq_3=2.34...$}
\vspace{0.5cm}

\begin{tabular}{|r|c|c|}\hline
N& D& E\\\hline
160&2.92863&6.08989\\\hline    
320&2.80693&5.27381\\\hline
640&3.05939&6.96764\\\hline 
\end{tabular}
\end{center}

As one observes the achieved accuracy does not exceed 15\%. Analysing
these numbers one can see that all the error comes from the region of
maximal $|n|$ and $|\nu|$ where the double Fourier transform is
apparently performed inaccurately.
>From these numbers we can only conclude that for this excited state
\beq
\epsilon\simeq 2.\pm 0.3
\eeq
In [1] the found value is 1.71231... Our result does not contradict
this number.\vspace{0.8cm}

{\bf 2.} We have also studied a possible solution for $q_3=0$, reported
in [1]. However in this case we were not able to construct a wave
function unique in the $r,\phi$ plane and satisfying the necessary
symmetry requirements. If, following [1] we seek $\Phi$ in the
form
\beq
\Phi(z,z^*)=\bar{u}Au
\eeq
where $u_i(z)$ are the three basic solutions of the eigenvalue equation
for $q_3$
and require the matrix $A$ to have certain matrix elements equal to zero
to make $\Phi$ finite, then at $q_3=0$ we find it impossible for $A$
to be invariant under the basic symmetry transformations of $z$. This 
seems to be related to the fact that at $q_3=0$ one may also take
$A_{33}\neq 0$. In any case there does not seem to exist solutions
for $A$ which preserve the required Bose symmetry.

Thus our conclusion is that the $q_3=0$ state reported in [1] does not
correspond to any physical odderon state.\vspace{0.8cm}

\section{"Moving" odderon}

For conformal weights $h=\frac{1}{2}+i\sigma$ the odderon energy
is supposed to behave at small $\sigma$ as
\beq
\epsilon(\sigma)=\epsilon_0+a\sigma^2
\eeq
where $\epsilon_0$ is the value (2) and $a$ is a parameter which
determines diffusion of the odderon wave function in the momentum space.
This parameter has been long known for the pomeron to be $14\zeta(3)$
(in units $3\alpha_s/\pi$). It is of certain interest to find $a$
for the odderon.

To this aim we first found the parameters of the odderon wave function
for various (small) $\sigma$ using the same method as employed for
$h=1/2$. Our results are presented in Table 3. The value of $iq_3$ turned
out to be real for arbitrary $\sigma$, whereas, with $\alpha$ chosen
to be real, both $\beta$ and $\gamma$ result complex. We chose $\alpha=1$

\begin{center}
{\bf Table 3. Odderon parameters for $h=\frac{1}{2}+i\sigma$}
\vspace{0.5cm}

\begin{tabular}{|r|c|c|c|}\hline
$\sigma$&$iq_3$&$\beta$&$\gamma$ \\\hline    
0.01&0.205306079&-0.971740164-i0.014404102&0.205305637-i0.000425478  \\\hline
0.1&0.210089247&-0.995153863-i0.142974530&0.210052319-i0.003938872    \\\hline
0.2&0.224303315&-1.060790013-i0.281079150&0.224222881-i0.006006327  \\\hline
0.3&0.247227544&-1.156524786-i0.415163678&0.247186043-i0.004529717  \\\hline 
0.5&0.316528176&-1.395571390-i0.695891904&0.316214188+i0.014095104    \\\hline
1.0&0.619239545&-2.044631201-i1.672240784&0.591391973+i0.183611401 \\\hline
\end{tabular}
\end{center}

Inspecting these figures one immediately notes that $|\gamma|=iq_3$.
This relation was predicted (for real $\gamma$) by L.N.Lipatov [7].

With the odderon parameters found we
calculated the odderon  energies directly,
using the same techique as for $\sigma=0$. With $\sigma$ different
from zero calculation becomes still more cumbersome and time and
memory consuming due to lack of certain symmetries and overall complex
arithmetics. For these reasons we had to limit ourselves with a
maximal 160 x 160 grid in the $r,\phi$ plane and neglected the
contribution from the asymptotical region $n>30$ $|\nu|>15$.
Our results are shown in Table 4 together with the ones just obtained
via the solution of the Baxter equation [8]

\begin{center}
{\bf Table 4. Odderon energies for $h=\frac{1}{2}+i\sigma$ }\vspace{0.5cm}

\begin{tabular}{|r|c|c|}\hline
$\sigma$&$\epsilon$ & $\epsilon$ [8]\\\hline
0.0&0.1534&0.16478\\\hline
0.1&0.1597&     \\\hline
0.3&0.2085&0.21777\\\hline
0.5&0.2980&0.30523\\\hline
1.0&0.6269&0.63228\\\hline
\end{tabular}
\end{center}

Our energies lie a little below the ones obtained from the Baxter equation,
which is natural since we have neglected the
asymptotic part  of the $n,\nu$ region in (4). Having this in mind we
find a complete agreement between our direct calculation results
and the ones based on the Baxter equation.

>From our energies  we find for the parameter $a$ in (18)
\[
  a=0.61
\]
More precise energies found in [8] lead to
\[
a=0.605
\]
Note however that already at $\sigma=1$ the approximation (18)
breaks down and more powers of $\sigma^2$ are needed to describe the
energy behaviour.
It is interesting that the parameter $a$ for the odderon is much smaller than
the one for the pomeron. In fact their ratio is of the same order as the
ratio of corresponding energies.

\section{Acknowledgements}
The author expresses his deep gratitude to Prof. J.Wosiek whose
comments initiated  this investigation and whose advice accompanied
its fulfillment. He is also grateful to Prof. L.N.Lipatov
for very interesting discussions. He is  grateful to both
of them for communicating their yet unpublished results.

\section{ References.}

1. R.A.Janik and J.Wosiek, Cracow preprint TPJU-2/98,
hep-th/9802100.

\noi 2. L.N.Lipatov, Phys. Lett. {\bf B309} (1993) 304.

\noi 3. R.A.Janik and J.Wosiek, Phys. Rev. Lett.{\bf 76} (1977) 2935;
R.A.Janik, Acta Phys. Polon. {\bf 27} (1996) 1819.

\noi 4. N.Armesto and M.A.Braun, Z.Phys., {\bf C63} (1997) 709.

\noi 5. P.Gauron, L.N.Lipatov and B.Nicolescu, Phys. Lett. {\bf B304} (1993)
334; Z.Phys. {\bf C63} (1994) 253.

\noi 6. M.A.Braun, St. Petersburg preprint SPbU-IP-1998/3, hep-ph/9801352.

\noi 7. L.N.Lipatov, {\it private communication}.

\noi 8. J.Wosiek,   {\it private communication}

\end{document}